\documentclass{aa}
\usepackage{graphicx}
\usepackage{natbib}
\bibpunct{(}{)}{;}{a}{}{,}

\newcommand\changed{}
\newcommand\changedd{}

\newlength{\picwidth}
\begin{document}

\title{Unusual Subpulse Modulation in PSR B0320+39}
\author{R.~T. Edwards\inst{1} 
   \and B.~W. Stappers\inst{1,2}
   \and A.~G.~J. van Leeuwen\inst{3}}
\institute{Astronomical Institute ``Anton Pannekoek'', 
        University of Amsterdam,
        Kruislaan 403, 1098 SJ Amsterdam, The Netherlands 
  \and
   Stichting ASTRON, Postbus 2, 7990 AA Dwingeloo, The Netherlands
  \and
   Astronomical Institute, Utrecht University, 
  Postbus 80000, 3508 TA Utrecht, The Netherlands
}
\offprints{R.~T. Edwards, \email{redwards@astro.uva.nl}}

\abstract{ 
We report on an analysis of the drifting subpulses of PSR B0320+39
that indicates a sudden step of $\sim180$ degrees in subpulse phase
near the centre of the pulse profile.  The phase step, in combination
with the attenuation of the periodic subpulse modulation at pulse
longitudes near the step, suggests that the patterns arise from the
addition of two superposed components of nearly opposite drift phase
and differing longitudinal dependence. We argue that since there
cannot be physical overlap of spark patterns on the polar cap, the
drift components must be associated with a kind of ``multiple
imaging'' of a single polar cap ``carousel'' spark pattern. One
possibility is that the two components correspond to refracted rays
originating from opposite sides of the polar cap. A second option
associates the components with emission from two altitudes in the
magnetosphere.
}

\maketitle
\section{Introduction}
For a number of pulsars it is known that each pulse is composed of a
number of subpulses, and that for each successive pulse the subpulses
appear to ``drift'' by a given amount across the profile \citep{dc68}.
The rate at which subpulses drift across the profile is not constant
but rather depends on their pulse longitude. This can be understood in
the context of the widely adopted ``carousel'' model of drifting
subpulses \citep{rud72}, which postulates the existence of a ring of
equally spaced ``sparks'' above the polar cap that give rise to
``tubes'' of plasma streaming upward from the surface, ducted along
magnetic field lines. At a certain altitude these particles emit
microwave radiation that is beamed along tangents to the local field
lines, giving rise to a beam pattern that consists of a ring of
``subbeams'' (bright spots), reflecting the polar cap spark
configuration. As the pulsar rotates the distant observer samples
emission along a certain path in the beam pattern. Under the right
viewing geometry, the sight-line makes a tangential pass along the
ring of subbeams, giving rise to the reception of a sequence of one
to a few subpulses every time the star rotates. Due to the non-linear
mapping between (spin) longitude and magnetic azimuth, the subpulses
should appear more closely spaced around the point at which the
sight-line makes its closest approach to the magnetic pole. As the
carousel slowly rotates about the magnetic pole, the longitude at
which the emission from a given spark is seen exhibits a monotonic
drift with time, which by extension of {\changed the} above argument
is greater in magnitude further from the magnetic pole.

By stacking the intensity time series modulo the pulse period to
produce a two-dimensional array in pulse number and pulse longitude,
and taking Fourier transforms along constant-longitude columns
(forming a longitude-resolved fluctuation spectrum; LRFS),
\citet{bac70b,bac70c} was able to measure not only the characteristic
periodicity in pulse number ($P_3$, the time taken for one subpulse to
drift to the former position of its neighbour), but also its amplitude and
phase as a function of pulse longitude. Many subsequent studies
employed this method but ignored the phase information, however those
that examined the longitude-phase relation
\citep{bac70b,wri81,dls+84,bmhm87,ash88} found behaviour qualitatively
consistent with that expected under the carousel model.

The longitude-phase relationship expected under the carousel model
depends on the number of sparks present, the misalignment angle
between the spin and magnetic axes, and their orientation with respect
to the line of sight. Hence studies of the relationship have the
potential to provide valuable information about the pulsar,
particularly in comparison to the observed longitudinal dependence of
the polarisation position angle which, under the magnetic pole model
of \citet{rc69a}, depends on the same geometric parameters.

Motivated by these goals, we embarked on a program of analysis of
several pulsars with drifting subpulses using the Westerbork Synthesis
Radio Telescope (WSRT). One of the target sources was PSR B0320+39.
This pulsar was discovered independently by \citet{dth78} and
\citet{iks82} in surveys of the northern sky at 400 and 102.5~MHz
respectively. Further observations revealed the emission of very
regular drifting subpulses \citep{iks82} which occur in two distinct
pulse longitude intervals, with a non-drifting component present in
the average profile \citep{ikl+93, kou00}. In this paper we report a
most unexpected result from our study of the longitude-phase
relationship of the drifting subpulses of PSR B0320+39.

\section{Observations and Analysis}
We observed PSR B0320+39 on 2000 November 3 with the Westerbork
Synthesis Radio Telescope (WSRT). Signals from fourteen telescopes were added
(after appropriate delays) in each sense of linear polarisation in a
10~MHz band centred at $328$~MHz. These were processed using the
Dutch Pulsar Machine (PuMa), operating as a digital filterbank
with 64 channels and $819.2$~$\mu$s sampling in total intensity
(Stokes I). For details of the system see \citet{vkv02}.

\begin{figure}
\resizebox{\picwidth}{!}{\includegraphics{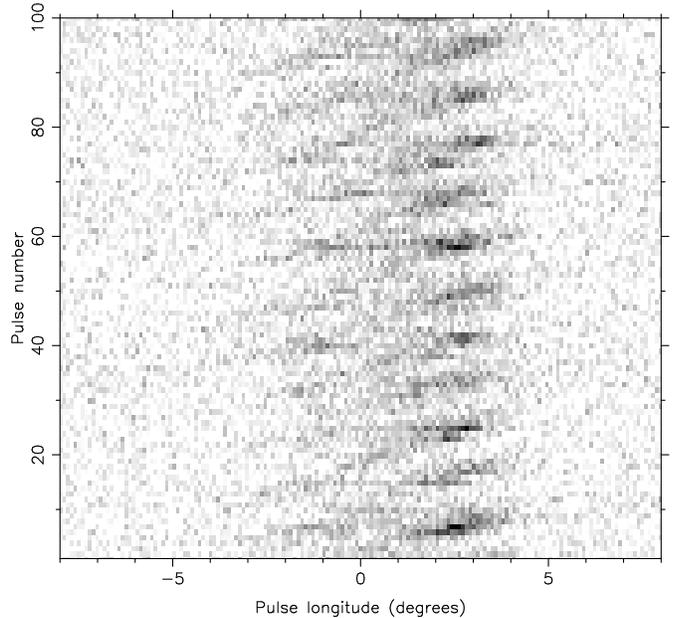}}
\caption{First 100 pulses of 328 MHz timeseries for PSR B0320+39,
binned in pulse longitude and pulse number.}
\label{fig:first100}
\end{figure}

In offline analysis we de-dispersed the data and
binned the resultant time series into an array in pulse
longitude and pulse number using the ephemeris of
\citet{antt94}. A total of 3400 pulses in 3701 longitude bins was recorded,
with a region of 512 bins including the on-pulse used in further
analysis. The samples were normalised to produce an average profile
with a zero baseline and a peak value of unity. The first 100 pulses
are shown in a greyscale form in Fig.~\ref{fig:first100}.

\begin{figure}
\resizebox{\picwidth}{!}{\includegraphics{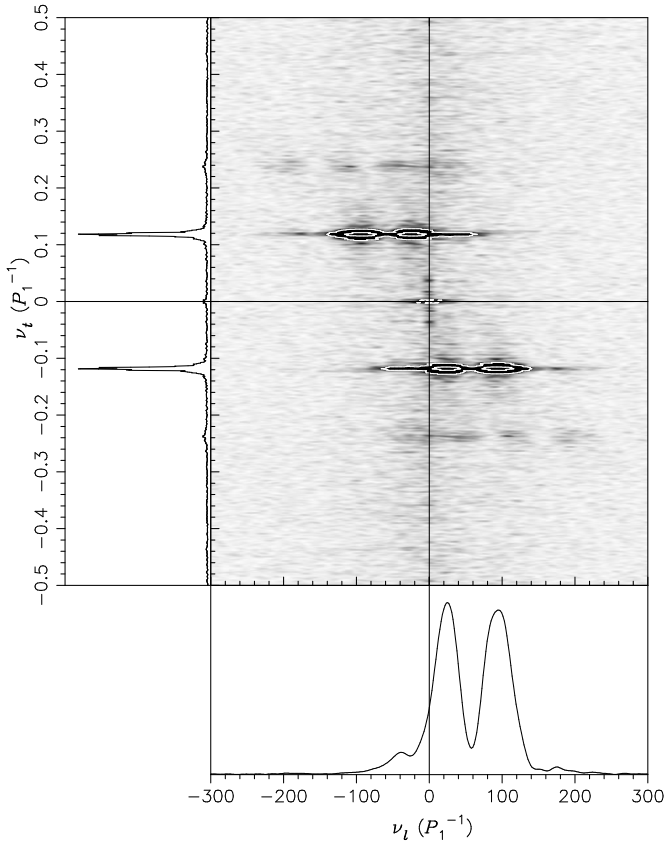}}
\caption{Main panel: Two-Dimensional Fluctuation Power Spectrum of PSR
B0320+39. The greyscale saturates at 0.05 times the peak power, while
the contours show the same spectrum at levels 0.05 and 0.5 times the
peak. Left panel: ``collapsed'' spectrum formed by averaging rows of
the 2D spectrum. Bottom panel: power as a function of $\nu_l$ for
$\nu_t=-0.118$.   }
\label{fig:spec}
\end{figure}

The techniques employed in the further analysis are based on those
presented in detail by \citet{es02}.  We computed the two-dimensional
Discrete Fourier Transform (DFT) of the first 2048 pulses of the pulse
longitude -- pulse number array to yield what we call the
two-dimensional fluctuation spectrum (2DFS). This is mathematically
equivalent to the harmonic resolved fluctuation spectrum (HRFS;
\citealt{dr99}; \citealt{es02})\footnote{\changed Without the use of
``twiddle factors'' \citep{ct65}, the 2DFS will differ from the
HRFS. However, it is easily shown that this merely amounts to the
provision of slightly different sampling lattices over the same
continuous function, due to the assignment of integer pulse numbers
(2DFS) versus integer harmonic numbers (HRFS).}, and the form of the
power spectrum found here for PSR B0320+39 matches that found by
\citet{kou00} for the HRFS. For the purposes of clear display we also
computed a spectrum using data from which we subtracted the average
profile from each pulse, and padded with zeros in the longitude axis,
in order to improve the resolution.  The resolution in the
time-associated axis was unnecessarily high (2048 elements), so to
improve the signal-to-noise ratio we smoothed the power spectrum by
convolution with a $7\times1$ -bin ($\nu_t \times \nu_l$) boxcar before
plotting the relevant portion of the spectrum\footnote{The sampling
interval used allows longitude-associated frequencies up to 1850
cycles/period, however all significant power is present in the portion
of the spectrum shown.}  in Fig.\ \ref{fig:spec}. The spectrum is
dominated by a pair of components associated with the drift
modulation; two are present because the spectrum derives from the
Fourier transform of real-valued data (see \citealt{es02}). Also
present with high significance are a pair of responses corresponding
to the second harmonic of the subpulse response. The features near
$\nu_t = 0.02$ and $0.04$ cycles/period are significant, however they also
appear in an analysis of off-pulse data, indicating that they are
caused by periodic interference.

After performing the 2D DFT, the complex spectrum was shifted in both
axes by an amount which placed the primary response to the drifting
subpulses at zero frequency. We then masked other components present
in the spectrum ({\changed namely the complex conjugate ``mirror'' of
the primary response, the pair of second harmonics and the DC
component}) {\changedd by multiplication of the coefficients with a transfer
function:
\begin{equation}
f(\nu_t) = \left\{      
              \begin{array}{ll}
                    0 &  |\nu_t-\nu_{tc}|/w < 0.5\\
                    2|\nu_t-\nu_{tc}|/w - 1 & 0.5 \leq |\nu_t-\nu_{tc}|/w \leq 1
\\
                    1 & |\nu_t-\nu_{tc}|/w > 1
              \end{array}
       \right. ,
\label{eq:filter}
\end{equation}
\noindent where $\nu_t$ is the time-associated axis, $\nu_{tc}$ is the
centre frequency of the component to mask, and $w$ is the frequency
extent of the region of zero transmission. The results that follow
derive from filters with $w=0.05$ cycles/period, although as expected
they were found to be insensitive to $w$ unless it becomes too small to
mask adequately or so large that it masks some of the desired signal.
We then computed the inverse DFT of the result.}
 
  We represent the subpulse signal as the real part of the product of
a pure two-dimensional sinusoid and a two-dimensional complex
modulation envelope.  By the convolution theorem, this envelope is the
result of the inverse transform performed above.  It describes the
deviations from a pure periodicity of constant amplitude that arise
due to amplitude modulation at the pulse period, scintillation or
nulling, non-uniform spacing of subpulses in pulse longitude (e.g. due
to sight-line curvature) and slow variations in the drift rate. Using
the iterative scheme of \citet{es02} we decomposed the two-dimensional
envelope into the product of two modulation envelopes that vary in
pulse longitude and pulse number respectively. Such a decomposition is
able to model the behaviour of a rotating carousel, where the geometry
defines the longitude-phase dependence and the carousel rotation
determines the time-phase dependence, and its basic assumptions --
that the longitudinal spacing of the subpulses is time invariant, and
that the time spacing of the subpulses is longitude independent -- are
in agreement with previous studies; see \citet{es02}.  The validity of
this approach in the case of PSR B0320+39 was {\changed confirmed} by
computation of the 2DFS of the difference between the {\changed
observed signal and that predicted using the real part of product of
the two one-dimensional envelopes.}
{\changed No significant power remained 
in the portion of the spectrum previously
occupied by the primary drifting subpulse response.}

The spectral shifting performed earlier has the effect of removing a
constant phase slope in each of the envelopes. Since the slope varies
over the envelope, the choice of value to use in the shifting is
somewhat arbitrary and was refined after viewing the resultant
envelopes. The periodicity with pulse number was well defined and a
value of $\hat{P_3}=-P_1/0.118$ was used. The response to the drifting
component in the other axis is very broad, and in fact extends well
into negative frequencies (Fig. \ref{fig:spec}). This complicates the
choice of value for $1/\hat{P_2}$ to be used in shifting the spectrum
in the longitude axis.  We chose a value of $\hat{P_2}=P_1/60$ to
give the resultant envelope a relatively small phase slope over most
of the pulse\footnote{\changed Since this value does not lie on the natural grid
of the DFT, the shift was actually performed
in the time-longitude domain by multiplication with a complex exponential.}.

\begin{figure}
\resizebox{\picwidth}{!}{\includegraphics{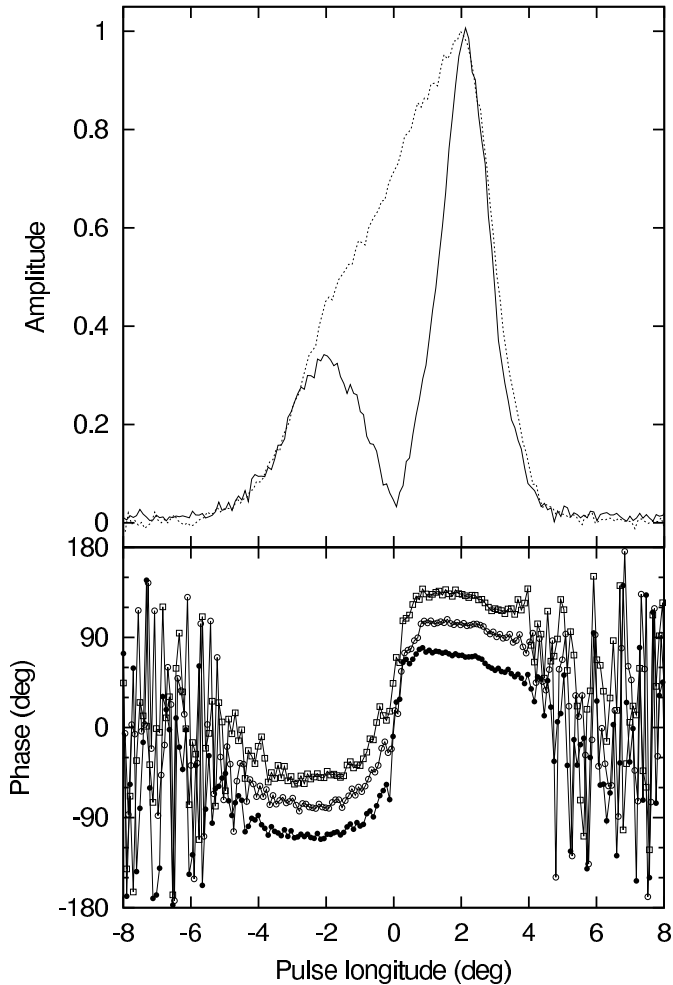}}
\caption{Inferred modulation envelope (solid line and filled circles)
and average pulse profile (dotted line). {\changedd Also shown is the phase
envelope computed by coherent addition of a set of 69 rows ($0.108 <
\nu_t < 0.128$) of the 3400-pt LRFS over which significant power was
seen (unfilled circles), and the phase envelope computed by coherent
addition of the peak row from each of twenty-six 128-pt LRF spectra
resulting from time-segmentation of the sequence (with this transform
length the majority of modulation power falls in a single row).}
The nominal phase slope of
$1/\hat{P_2}=60\degr/\degr$ has been subtracted from the phase envelopes
before plotting.}
\label{fig:env}
\end{figure}

\begin{figure}
\resizebox{\picwidth}{!}{\includegraphics{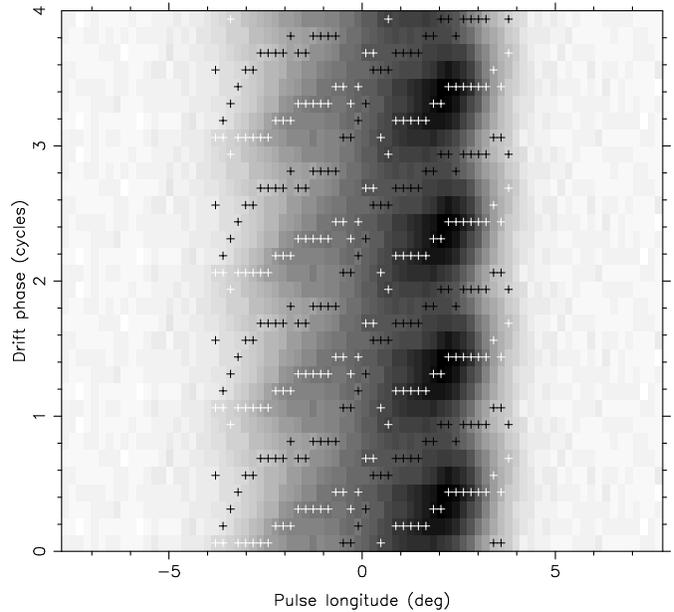}}
\caption{Data folded at $P_3$, as a function of pulse longitude and
drift phase. Maxima and minima in each longitude bin are indicated
with black and white crosses respectively. The data are plotted four
times to guide the eye.}
\label{fig:folded}
\end{figure}

The inferred longitude-dependent envelope (Fig. \ref{fig:env}) shows
that nearly linear subpulse modulation occurs in two distinct
longitude intervals, with a striking {\changed phase offset} of
$\sim$180\degr\ between them.  In terms of the longitude-time diagram,
the effect of the {\changed offset} is to cause the drift band peaks in one
component to extrapolate to troughs in the other and vice versa. To check
this result we computed the (complex) LRFS and examined the amplitudes
and phases in frequency bins near $\hat{P_3}=P_1/0.118$.  {\changed
Since the modulation is not a pure periodicity,
the power is spread over several bins in the spectrum. Although of a
much lower signal-to-noise ratio, the shape of the phase envelope in
each bin was consistent with that inferred with our preferred
technique, which makes optimal use of all modulation power. Similarly,
division of the observation into segments in time (which results in
wider frequency bins and potentially also in reduced frequency drift
in each segment) and examination of the phase relation in the peak bin
of each segment also gave consistent results (regardless of the
segment size). In order to improve the signal-to-noise ratio of these
other methods, we also used an iterative algorithm based on
multiplication with the complex conjugate of a template envelope to
determine the correct phases with which to add envelopes from
different frequency bins or time segments (producing in effect
coarse-grained versions of the algorithm of \citealt{es02}). The
resultant summed envelopes were again consistent with the result of
our preferred algorithm {\changedd (Fig. \ref{fig:env})}. } As a final
check, we folded the longitude-time data about the pulse number axis
(modulo $1/0.118$ pulses) to produce an average profile as a function
of drift phase and pulse longitude (Fig. \ref{fig:folded}). As with
the LRFS methods, much of the modulation was lost due to the unstable
drift motion, however some residual modulation remained and is
sufficient to confirm the unusual phase behaviour.

Knowing that there is a $\sim$180\degr\ {\changed offset between the
two halves of the} subpulse phase envelope, we may explain two
otherwise curious facts about PSR B0320+39.  The first concerns the
values of $P_2$ previously reported, which are around $P_1/120$
(\citealt{ikl+93} report $3.1\degr$ of longitude, i.e. $P_2 \simeq
P_1/116$ at 406~MHz; \citealt{kou00} reports $24.5$~ms, i.e. $P_2
\simeq 123$, at 328~MHz).  These strongly differ from our nominal
$\hat{P_2}=P_1/60$ because they are based on measurements of actual
subpulse separations, which will nearly always be taken between
subpulses on opposite sides of the phase {\changed shift}. Since the
subpulse in the trailing component leads (or lags) the position it
would take in a continuous linear phase envelope by $\sim 180\degr$ of
subpulse phase, the derived spacing differs from that expected from
the typical phase slope by a factor of 1/2 (and probably sometimes
3/2).
The second curiosity explained by the phase {\changed offset} is the
local minimum in the subpulse HRFS response at around 60 cycles/period
(Fig.\ \ref{fig:spec}).  The complex longitude-dependent modulation
envelope (once the constant phase slope is subtracted) is quite close
to an odd real-valued function, which causes its Fourier transform to
have local minimum in its power at DC. This is the cause of the local
minimum around $60$ cycles/period, the typical longitudinal drift
frequency. Without the phase {\changed offset} the envelope would be close to a
real-valued even function and its Fourier transform would not have a
strong minimum at its centre. 

Furthermore, we note that the equivalence of the 2DFS and HRFS
provides an explanation for the apparent ``secondary'' component at
$0.118$ cycles/period, suggested by \citet{kou00} to be due to the
ambiguity between which drift band joins to which between the two
components. In fact, the subpulse response is very broad in the
longitude-associated frequency axis, and extends significantly into
negative frequencies. The symmetry of the two dimensional Fourier
transform of real functions is such that the power at any frequency
pair $\nu_l, \nu_t$ is equal to that at $-\nu_l, -\nu_t$ (where
$\nu_l$ and $\nu_t$ are the frequencies in the longitude-associated
and time-associated axes respectively). If only the positive
frequencies in $\nu_l$ are plotted, as is the case with the HRFS, no
information is lost but individual components that span the $\nu_l=0$
axis will appear as two spatially separated components at $\nu_t$
values of opposite sign. When the full spectrum is plotted as in
Fig.\ \ref{fig:spec} the proper interpretation becomes apparent.

\begin{figure}
\resizebox{\picwidth}{!}{\includegraphics{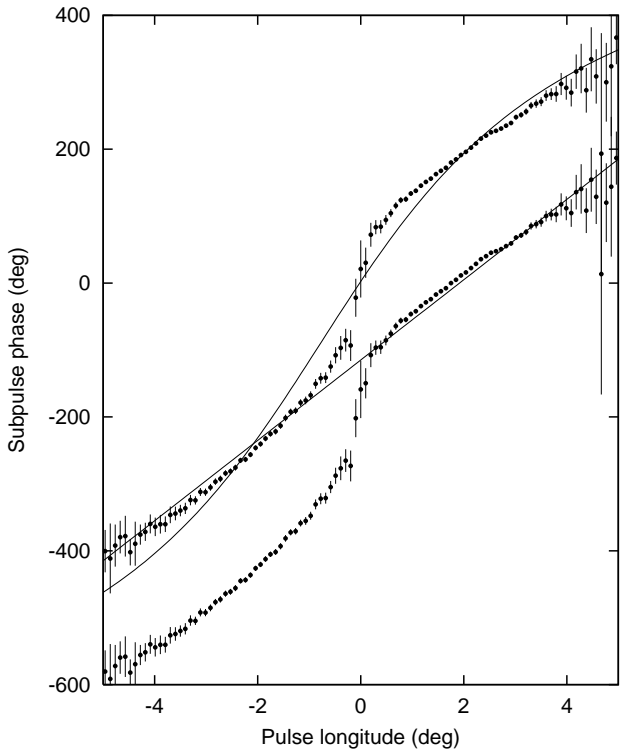}}
\caption{\changed Inferred subpulse phase envelope, including the
$60\degr/\degr$ phase slope previously removed (indicated by the straight
line). The data are plotted twice with
an offset of $180\degr$ to facilate evaluation of the central feature
as the transition between two offset regions of linear
drift. Two-sigma error bars are included. The curved line represents
the best fit to the data under the standard carousel model, given the
polarimetric constraint (see text). }
\label{fig:phase}
\end{figure}

\section{Discussion \& Interpretation}
\subsection{Polar Cap Models}
The phenomenon of drifting subpulses has been used repeatedly in
investigations of pulsar emission as a potentially strong clue about
the geometry of the system.  The fact that those pulsars that show
nearly linear drifting subpulses tend to have a double, ``unresolved
double'' or ``single'' average pulse profile morphologies, with many
showing a transition from resolved to unresolved double as the
observing frequency is increased, is usually seen as strong support
for some form of ``hollow-cone'' polar cap model
\citep{rc69a,rs75,bac76,ran83,ran86,lm88}. PSR B0320+39 shares these
characteristics of regular drifting subpulses and a transition from a
resolved double to an unresolved double average profile morphology
between 102~MHz \citep{iks82}\footnote{Due to an improved system
response the profile of \citet{kl99c}  shows much stronger component
separation than that of \citet{iks82}; see fig.~\ref{fig:profs}.} and
328~MHz (this work; also \citealt{dth78,ikl+93}).

{\changed The local steepening of the phase envelope
of PSR B0320+39 in the transition region is inconsistent 
with the carousel model in its most basic form.}  For a single ring of
``sparks'', as long as the sparks are elongated in either magnetic
azimuth or opening angle (or neither, but not both), the subpulse
modulation seen in a ring of constant opening angle has the same phase
as that in any other ring, if measured from the same fiducial
azimuth. Therefore, the observed longitude-dependent subpulse phase
envelope is directly tied to to the magnetic azimuth sampling of the
sight-line (via multiplication by the number of sparks, $N$)
\citep{es02}.  The only way to produce a {\changed sudden swing} in magnetic
azimuth (and hence subpulse phase) is to pass close to the magnetic
pole. {\changed Such a configuration is inconsistent with the usual
interpretation of unresolved double profile morphologies, and with the
gradual linear polarisation position angle swing observed in this
pulsar \citep{sp02}. To prove this assertion we attempted to fit the
observed subpulse phase swing with the geometric formula of
\citet{es02}, using a $\chi^2$ minimisation algorithm. From the figure
provided by \citet{sp02} we conservatively estimate the range of
polarization position angle gradients consistent with the data as
${\mathrm d}\chi/{\mathrm d}\phi = 12 \pm 3 \degr/\degr$ (where $\chi$
and $\phi$ are the position angle and pulse longitude
respectively). The parameters of the fit were required to be
consistent with this gradient via the geometric model of
\citet{rc69a}.  The result (Fig.~\ref{fig:phase}) clearly fails to
describe the observed behaviour.}

One way to produce a jump in subpulse phase is to adopt
a nested ring polar cap spark configuration \citep{gs00} and to assign
the two halves of the profile to rings of different opening angles and
numbers of sparks. However, this would result in a measurably
different $P_3$ unless the circulation rates of the two carousels were
different by a fortuitous ratio (i.e. the inverse of the ratio of the
numbers of sparks in the rings). In addition, the transition from
resolved to unresolved double profile with increasing frequency could
not be explained by standard radius-to-frequency mapping (RFM;
\citealt{cor78}). Some other mechanism such as the finite beam-width
of the elementary emission mechanism ($2\gamma^{-1}$), the finite
band-width of emission at a given radius (with attendant RFM in the
effective range of radii visible), or refractive broadening would be
required to produce the profile evolution.

\subsection{Double Imaging of Magnetospheric Origin}
\subsubsection{Basic Considerations}
{\changed Given the failure of the standard model to account for the
observed subpulse drift behaviour of PSR B0320+39, we develop in this
section a pair of explanations based on a simple and plausible
extension of the model. Namely, we argue that the most important
features of the longitude-dependent subpulse modulation envelope -- a
local reduction of amplitude accompanied by a rapid swing in the phase
angle -- are the result of destructive interference between two
superposed drifting subpulse signals of nearly opposite phase. The
observed behaviour and this explanation of it can be understood by
analogy with the phenomenon of orthogonal polarisation mode
transitions, where the pulsar signal is seen to consist of the
incoherent sum of two signals with nearly orthogonal polarisation
states (e.g.~\citealt{mth75,scr+84}). In both cases the total
intensity of the two signals simply add in the average profile, while
the near anti-parallel nature of the vectors representing the complex
subpulse modulation signal (in the Argand plane) or the polarised part
of the signal (in the Poincar\'e sphere) explains the reduction in
magnitude (subpulse amplitude, polarised intensity) and the swing in
orientation (subpulse phase, polarisation position angle) during a
transition of dominance from one mode to the other. Other than the
obvious features of the shift in phase and the attenuation in
amplitude, the continuity of both the average profile and the absolute
value of the gradient of the subpulse amplitude envelope serve as
supporting evidence for this hypothesis.}

Since the polar cap cannot
be physically populated with overlapping spark systems, we argue that
the superposed drifting components of PSR B0320+39 are associated with
a form of ``double imaging'' of the polar cap excitation
pattern. Whereas the polar cap may consist of a single, drifting ring
of excitation points (e.g. sparks), the physics that map this to a
radiated beam pattern must do so under two different transformations
that are received in incoherent addition by the observer. This gives
rise to a subpulse pattern that is the sum of two overlapping
components with differing peak pulse longitudes, and phases that
differ in their contribution at a given pulse longitude by a constant
offset of $\sim 180\degr$. This also implies that the average pulse
profile is the simple sum of two overlapping components approximately
equal to the corresponding subpulse amplitude envelopes.

\begin{figure}
\resizebox{\picwidth}{!}{\includegraphics{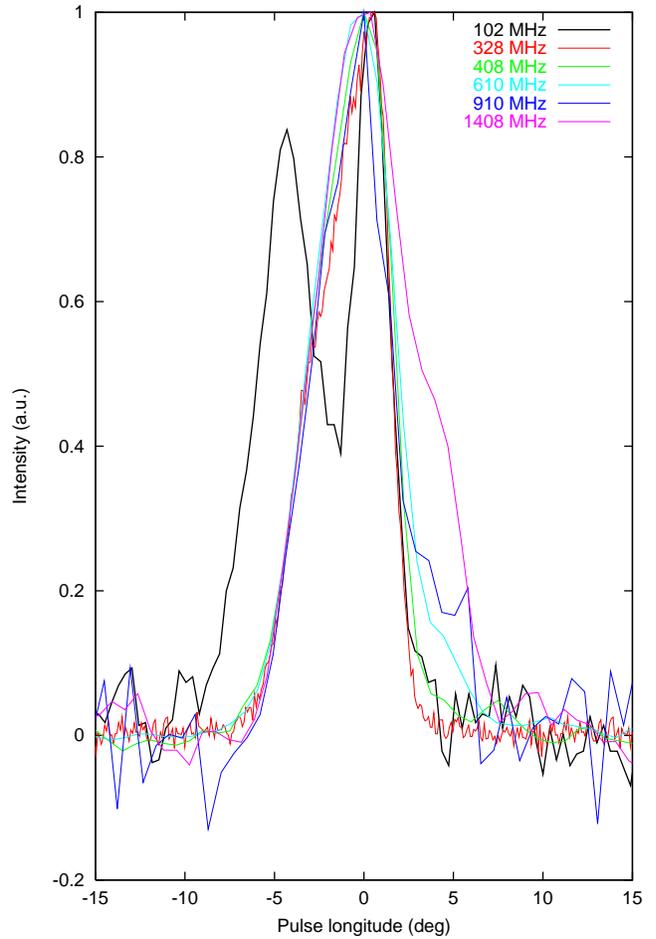}}
\caption{Average total intensity pulse profiles of PSR B0320+39 over a
frequency range of 102--1408 MHz. The profiles are normalised to the
same peak intensity and (in the absence of timing information) were
aligned by eye to match the model proposed here. The 328~MHz profile
is from this work. The 102~MHz profile is that of \citet{kl99c}, and
the remaining profiles are from \citet{gl98}. All profiles other than
the 328~MHz one were obtained via the EPN profile database.}
\label{fig:profs}
\end{figure}

In this section we discuss two potential mechanisms for such an
effect. The first associates the images with different ray paths in a
refractive magnetosphere \citep{pl00}, while the second attributes
them to emission originating at two discrete altitudes in the
magnetosphere \citep{ran93}. Before going into the details of each
scenario, given that they both produce a beam consisting of a pair of
nested cones, we are are in a position to examine the consistency
between these models and the observed pulse profile morphology. This
is important because, in assigning the two components of PSR B0320+39
to cones of different opening angles, we are abandoning the standard
``single cone'' explanation favoured for most pulsars with regular
drifting subpulses. We suggest instead that a cut through an outer
cone (from the outside to the inside) produces a leading component,
while a trailing component is produced by a pass along the edge of the
inner cone. At most radio frequencies the difference in the opening
angles of the cones is small and the components overlap, but
radius-to-frequency mapping causes the outer cone to expand at lower
frequencies, giving the components sufficient separation to be
resolved at 100~MHz. Clearly one should also expect a third component
to be present on the trailing edge of the profile as the sight-line
passes across the outer cone for the second time. This is actually
observed at frequencies at and above 600~MHz \citep{gl98}, giving
further support to our assertion that more than one cone of emission
is visible from this pulsar. As shown in Fig.\ref{fig:profs}, the
pulse profiles observed for PSR B0320+39 match this picture well.  We
now turn to the specifics of the two mechanisms.

\begin{figure}[t]
\centerline{\resizebox{0.6\picwidth}{!}{\includegraphics{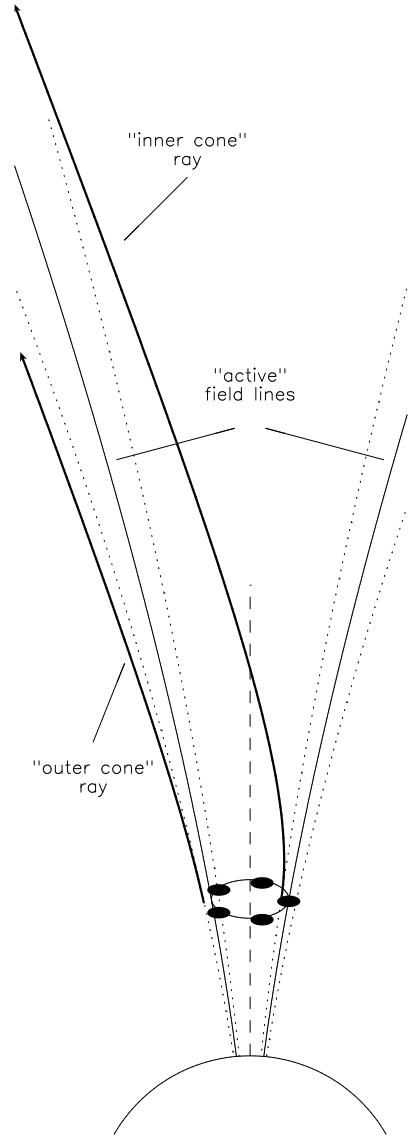}}}
\caption{Schematic illustration of the simultaneous observability of
rays from opposite sides of the polar cap, via magnetospheric
refraction. The plane of the page contains the magnetic axis, the
observer, and all possible origins for observable rays emitted and
refracted under axisymmetric conditions.  The ``active'' field lines
that mark the peak of the plasma distribution at a given altitude are
labelled. The field lines defining the approximate edge of the
emitting flux tubes are shown with dotted lines. Two rays,
corresponding to those seen by an observer whose sightline presently
intersects the overlap region of two conal beams, are shown.  The left
hand ray originates outside the peak of the plasma distribution, and
is refracted further in this direction. The right hand ray originates
inward of the peak plasma density and is refracted across the pole to
exit at the same angle as the other ray. A carousel-like intensity
pattern with 5 subbeams is shown to illustrate the origin of the
subpulse phase difference between the two components associated with
these ray paths.}
\label{fig:refract}
\end{figure}
\begin{figure}[t]
\centerline{\resizebox{0.6\picwidth}{!}{\includegraphics{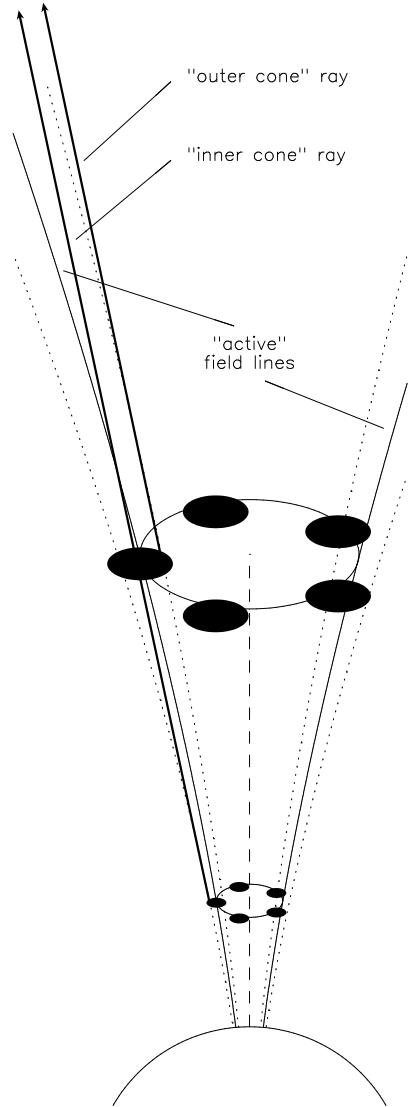}}}
\caption{Schematic illustration of the simultaneous observability of
rays from two different emission heights (see also
Fig. \ref{fig:refract}). Due to the curvature of the magnetic field lines
along which the radiating particles stream, the ray from the upper
region must come from a more inward field line that that from the
lower region, if the two are to make the same angle to the magnetic
axis.}
\label{fig:dc}
\end{figure}

\subsubsection{Refraction}
Refractive effects have long been considered potentially important in
pulsar magnetospheres (e.g. \citealt{mel79,ba86}), particularly in the
production of circular polarisation \citep{am82} and superposed
orthogonally polarised radiation \citep{mck92}. Only recently, however,
has the potential for refraction to explain pulse profile morphologies
seen detailed study \citep{lp98,pl00,pet00}.  Based on the simple
argument that the magnetospheric plasma density has minima both at the
magnetic pole and at the edges of the cone defined by the last open
magnetic field lines, \citet{pl00} showed that rays emitted poleward
of the peak of the plasma distribution could be refracted {\it across}
the pole to exit with an angle to the pole of the opposite sign to
that with which they were emitted. Rays emitted outward of the plasma
peak are simply refracted to somewhat greater opening angles.  This is
illustrated schematically in Fig. \ref{fig:refract}. Clearly, at any
pulse longitude the observer may potentially receive two rays, one
from either side of the magnetic pole.

In an axisymmetric plasma distribution, refraction occurs purely in
the plane of the magnetic field lines. \cite{pet00} shows that the
magnetosphere tends to concentrate rays from within a given set of
co-planar field lines into two components of different peak opening
angle, corresponding to the different ray paths described above. These
are associated with inner and outer components of the conal beam,
which in terms of explaining the morphology of pulsar average profiles
serves the same purpose as the double cone of \citet{ran93} (see
below). This grouping of rays occurs within a given plane of field
lines; all possible origins for visible rays must be co-planar, along
with the line of sight and the magnetic axis. Therefore, to the extent
that the plasma distribution is axisymmetric, the geometric
conveniences of the basic magnetic pole model \citep{rc69a} are
preserved, including the polarisation behaviour, and the mapping of
azimuthal intensity modulations (e.g. from a plasma flow initiated by
a polar cap spark carousel) to observable drifting
subpulses.

 Moreover, if the polar cap pattern consists of a ring of an {\it odd}
number of points of excitation, the subpulse pattern seen in the
component associated with pole-crossing rays should be close to
$180\degr$ out of phase with that in the other component. 
{\changed The slight deviation from purely out of phase signals required to
explain the finite longitude interval over which the phase transition
occurs} might be explained as a retardation effect or a refractive delay.
Alternatively, inclusion of magnetospheric rotation
effects or refraction under a moderately non-axissymmetric plasma
distribution (\citealt{pl00} model both but only in the context of
polarisation wave mode coupling; see also \citealt{bcw91}) could
potentially provide the kind of symmetry-breaking needed to produce
subpulse phase offsets other than 0 or 180\degr. The form of the beam
pattern and the manner in which a rotating carousel polar cap system
maps to drifting subbeams under such conditions is yet to be
investigated, to our knowledge.

We note that this model can also explain a recent result concerning
the drifting subpulses of PSR B0809+74. \cite{rrs+02} found that over
the course of each subpulse, the polarisation state jumps in a regular
manner from one polarisation state to another, almost orthogonal
state. This is similar to the behaviour seen in PSR B0320+39, in that
the observed total intensity drifting subpulse pattern can be
understood as the sum of two components with a phase offset between
them. However, in PSR B0320+39 the drift components cannot be
associated with orthogonal polarisation modes since the linearly
polarised average profile at 408~MHz peaks in intensity at the centre
of the profile \citep{gl98}, where maximal depolarisation ought to
take place, were the components orthogonally polarised. The
single-pulse polarimetric work of \citet{sp02} confirms that the
situation is similar at 103~MHz: significant power is present in two
modes only in the profile wings. The model of \citet{pet01} for the
origin of orthogonal polarisation modes offers a simple explanation
for this disparity. Under that model the secondary polarisation mode
is produced via the conversion of power from the ordinary mode in
regions of propagation that are quasi-longitudinal to the magnetic
field. If radiation in one of the conal components experiences
significant conversion while the other does not, as could well occur
since the ray trajectories involved are very different between the
two, the situation is as seen in PSR B0809+74. If the configuration is
not conducive to conversion then both components have similar
polarisation, as in PSR B0320+39.  Since the conversion process also
alters the orientation (i.e. position angle, ellipticity) of the
modes, the longitude-dependent non-orthogonality of the polarisation
modes reported by \cite{rrs+02} might also receive an explanation via
this effect.

\subsubsection{Dual Emission Altitudes}
A second potential explanation for how we receive the sum of two
related subpulse signals is that the two signals arise at different
altitudes in the magnetosphere.  This is the essence of the ``double
conal'' model of \citet{ran93}, so-called because the production of
radiation cones at two distinct altitudes in the magnetosphere gives
rise to a beam pattern that consists of two nested cones. The radial
separation of the emitting regions and the differential tangent angle
between the active field lines and the magnetic axis that arises as a
result causes the polar cap plasma flux cone to produce a wider
radiation beam at the upper emission region, compared to the lower
one. If the height difference is large, up to four well-resolved conal
profile components are produced, while reduced vertical separation
causes the components to overlap and produce a ``boxy'' profile
shape. Fig. \ref{fig:dc} illustrates this model as it might apply for
PSR B0320+39 at 102~MHz, with the conal beams overlapping to a lesser
degree than at higher frequencies.

Clearly this model is able to explain how we see superposed drifting
patterns, but we have not yet offered an explanation for why there
should be a subpulse phase offset between them. They arise from the
same pattern of sparks (or plasma flux production) on the polar cap,
yet we observe a phase difference of $\sim180\degr$ near the
profile centre.
For the subpulse pattern to be delayed by $P_2/2 \simeq 30$~ms due to
simple retardation \citep{cor78} requires that the emission regions
are separated by $\sim 9000$~km, 1--2 orders of magnitude greater than
the values typically inferred from profile morphologies
(e.g. \citealt{gg01,mr02a,gg03}) . Including differential aberration
reduces this by no more than a factor of 2 \citep{cor78}. Hence we
consider this explanation unlikely. An alternative explanation worth
pursuing in the future, but outside the scope of this work, is that a
significant electric field exists throughout the magnetosphere,
causing particles to accelerate upward mostly along magnetic field
lines, but with some azimuthal $\vec{E}\times\vec{B}$ drift
(e.g. \citealt{wri02}). The drift would cause the beam pattern
radiated at a given height to be rotated in magnetic azimuth by a
certain amount, which may account for the subpulse phase offset seen
here.

In any case, under this model the phase difference must ultimately be
due to the different altitudes at which the components originate.
Since RFM must apply to the outer (i.e. upper)
component to explain the 102~MHz pulse profile, the difference in
subpulse phase between the components ought to be a strong
function of frequency. Although this is the first study to examine the
longitude dependence of subpulse phase in this pulsar, the fact that
the longitude-dependent subpulse amplitude envelopes measured at
102~MHz and 408 MHz by \citet{ikl+93} drop to close to zero in the
profile centre, just as at 328~MHz, suggests that at all three
frequencies the subpulses add destructively in this interval, and
therefore that the phase difference between them is close to
$180\degr$, regardless of frequency. The origin of the phase {\changed offset}
therefore remains a mystery under this scenario.

\section{Conclusions}
The drifting subpulses of PSR B0320+39 occur in two distinct intervals
of pulse longitude. Using a sensitive new technique we have shown that
the extrapolation of subpulse drift from a given longitude interval is
$\sim 180\degr$ out of phase with the subpulses of the
other. This was confirmed by analysis of the longitude-resolved
fluctuation spectrum and by folding the data into profiles
corresponding to different phases of the subpulse drift.

This suggests the presence of two components of nearly opposite drift
phase, and based on the forms of the average pulse profile and the
longitudinal dependence of the amplitude of subpulses, we argued that
the modes are superposed at all longitudes. The adding of intensity
contributions from the two modes results in attenuation of the
modulations in the longitude interval where the modes are of similar
intensity, and produces a steep {\changed transition of $\sim
180\degr$} in the longitudinal subpulse phase envelope at the point
where the component intensities are equal.

Two specific models are suggested to explain this behaviour. Both
produce a nested pair of conal beams, one via magnetospheric
refraction, the other via emission at two distinct
altitudes. The leading and trailing components are
associated with outer and inner cones respectively. The second cut of
the outer cone does not give rise to a component at $102$~MHz or
$328$~MHz, but does produce a third, trailing component at frequencies
above $600$~MHz. The component separation increases rapidly at
frequencies below $328$~MHz, in accordance with the expectations of
radius to frequency mapping. The refractive scenario can explain the
observed phase {\changed offset} as a simple consequence of having an
odd number of subbeams in the carousel, whilst under the dual-altitude
emission model the phase {\changed offset} is more difficult to understand.
Planned observations of the time and frequency dependence of the phase
offset, and the behaviour of subpulses in the third component present
at and above $600$~MHz, should prove most helpful in the further
evaluation of these models.

\acknowledgements The authors wish to thank P.~Weltevrede for
enlightening discussions, and the referee whose persistent skepticism
led to the correction of an error in our analysis. RTE is
supported by a NOVA fellowship.  BWS is supported by NWO Spinoza grant
08-0 to E.~P.~J.~van den Heuvel.  The Westerbork Synthesis Radio
Telescope is administered by ASTRON with support from the Netherlands
Foundation For Radio Astronomy.  Part of this research has made use of
the data base of published pulse profiles maintained by the European
Pulsar Network, available at
http://www.mpifr-bonn.mpg.de/pulsar/data/.

%\bibliographystyle{aa}
%\bibliography{journals_apj,local,modrefs,psrrefs,grbrefs,crossrefs}

\end{document}